\documentclass[aps,preprint,showpacs,groupedaddress]{revtex4}  
\usepackage{graphicx}  
\usepackage{dcolumn}   
\usepackage{bm}        
\usepackage{amssymb}   

\begin{document}
\title{Collisionless Magnetic Reconnection via Alfv{\'e}n Eigenmodes}
\author{ Lei \surname{Dai}}
\affiliation{School of Physics and Astronomy, University of Minnesota, Minneapolis, Minnesota 55455, USA}
\date{\today}

\begin{abstract}
We propose an analytic approach to the problem of collisionless magnetic reconnection formulated as a process of Alfv{\'e}n eigenmodes' generation and dissipation. Alfv{\'e}n eigenmodes are confined by the current sheet in the same way that quantum mechanical waves are confined by the $\tanh^2$ potential. The dynamical timescale of reconnection is the system scale divided by the eigenvalue propagation velocity of the n=1 mode. The prediction of the n=1 mode shows good agreement with the \textit{in situ} measurement of the reconnection-associated Hall fields. 
\end{abstract}

\maketitle 
Magnetic reconnection is recognized as a universal process that converts magnetic field energy to the kinetic and thermal energy of the plasma in space and laboratory. The physical mechanism of energy conversion and the time scale are the fundamental issues of reconnection. Energy conversion in the Sweet-Parker model \cite{Sweet1958,Parker1963} is done by means of magnetic field diffusion into the plasma fluid with the characteristic velocity being the diffusion velocity. Petschek introduced a mechanism as a complement to diffusion where the change in magnetic field propagates as a slow mode shock \cite{Petschek1964}. The characteristic velocity within Petschek model is the shock propagation velocity, which can be much larger than the Sweet-Parker's diffusion velocity. In these models the time scale of reconnection is estimated as the system scale divided by the characteristic velocity; the reconnection rate is measured by the plasma inflow velocity balanced by the characteristic velocity in a steady state. \\

However, crucial aspects of reconnection in real plasmas such as onset and temporal behavior cannot be resolved in steady theories. The general approach to unsteady reconnection has been to Fourier analyze current sheet dynamics in $(\omega, \vec{k})$ space and search for instabilities, e.g. the tearing mode \cite{Furth1963}. Unstable reconnection modes grow significantly on time scale measured by the growth rate. In this letter we present a novel approach to time-dependent collisionless reconnection. Collisionless reconnection is described as the generation and dissipation of Alfv{\'e}n eigenmodes. Not only can Alfv{\'e}n eigenmodes grow, they can also be damped by transferring wave energy into reconnection ion jets. We solve linearized two-fluid equations to find the self-consistent evolution of reconnection following an initial perturbation in similar spirit of Landau's method \cite{Landau1946, Stix1992}. The dynamical timescale of reconnection is the system scale divided by the eigenvalue propagation velocity of the n=1 mode. Both Fourier analysis of instabilities and this theory are limited in the linear regime. \\

\begin{figure}
\scalebox{1.00}{\includegraphics{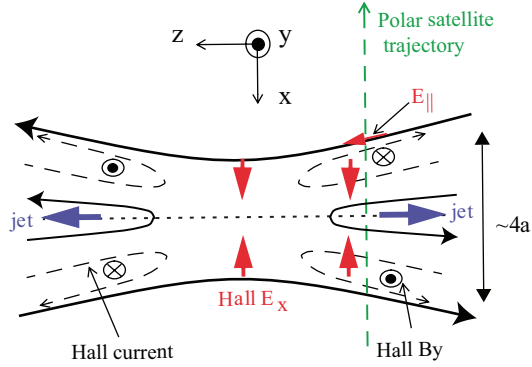}}
\caption{Schematic diagram of Magnetic Reconnection in the current sheet with Hall fields and current included. The Hall fields, the Hall current and the magnetic-field-aligned electric field are incorporated into the Alfven eigenmode. Predictions of the Alfv{\'e}n eigenmode are compared with measurements from the Polar satellite in Fig.~\ref{fig:fig2}. Analysis suggests a phase with opposite Hall current and Hall $B_y$ preceding the phase in 
figure.}
\label{fig:fig1}
\end{figure}

\begin{figure*}
\scalebox{.90}{\includegraphics{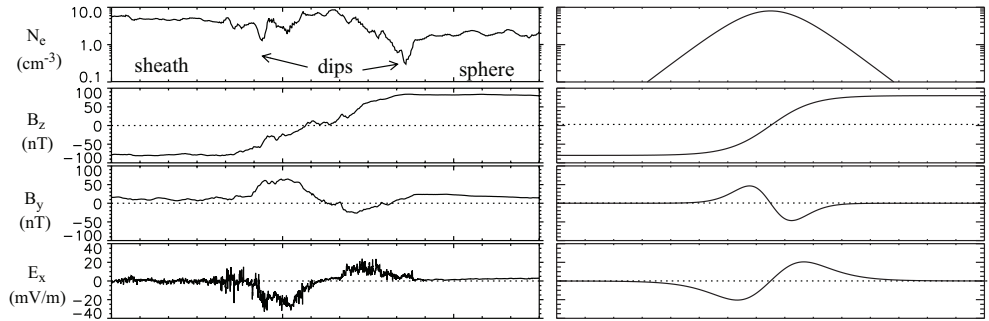}}
\caption{Comparison between the observation by Polar (left column) and predictions of the n=1 Alfv{\'e}n eigenmode in the Harris sheet (right column). The figure of measurements is from Ref.\cite{Mozer2002}. Unit length in $x$ is 100km.} 
\label{fig:fig2}
\end{figure*}

The experimental motivation for this Alfv{\'e}n-eigenmode approach is the recent \textit{in situ} measurement of reconnection-associated Hall fields in collisionless space plasmas \cite{Mozer2002,Wygant2005,Vaivads2004,Oieroset2001} (see Fig.\ref{fig:fig1}) and laboratory plasmas \cite{Ren2005,Yamada2006}. Hall fields and current were first introduced by Sonnerup \cite{Sonnerup1979} as a steady structure in the diffusion region. Later numerical studies looked at other various perspectives \cite{Terasawa1983,GEM2001}. In our approach, Hall fields and current are incorporated into the Alfv{\'e}n eigenmode. Predictions of the n=1 mode show good agreement with \textit{in situ} measurements (see Fig.\ref{fig:fig2}). Hall perturbations are smaller than or, at most, comparable to the background as indicated by measurements \cite{Mozer2002,Wygant2005,Vaivads2004,Oieroset2001,Ren2005,Yamada2006}. This fact implies that a linear theory may suffice to explain the essential physics of collisionless reconnection.\\ 

The set of collisionless two-fluid equations is 
\begin{eqnarray}
c\nabla \times\mathbf{B}=4\pi(\mathbf{J_i+J_e}),\label{eq:Ampere}
\end{eqnarray}
\begin{eqnarray}
c\nabla \times\mathbf{E}=-\partial_t \mathbf{B},\label{eq:Faraday}
\end{eqnarray}
\begin{eqnarray}
\partial_t(n_sq_s)+\nabla \cdot\mathbf{J_s}=0,\label{eq:continuity}
\end{eqnarray}
\begin{eqnarray}
m_sn_s\partial_t\mathbf{u_s}+m_sn_s(\mathbf{u_s}\cdot\nabla) \mathbf{u_s}=n_s q_s(\mathbf{E}+\frac{\mathbf{u_s}}{c}\times \mathbf{B})-\nabla n_sT_s,\label{eq:momentum}
\end{eqnarray}
where the subscript $s$ represents the particle species. Ions and electrons have been assumed isotropic and isothermal. We also assume a quasi neutral plasma, $n_i\approx n_e$. The coordinate system is depicted in Fig.~\ref{fig:fig1}, with x along the normal of the current sheet, z parallel to the background magnetic field, and y aligned with the background current. $\partial_y=0$ is assumed. Initially the current sheet is set as a Harris sheet \cite{Harris1962} with background plasma density $n_s=n_0{\rm sech}^2(x/a)$, background magnetic field $B_z=B_o\tanh (-x/a)$ and the background current $J_{yo}$ supported by the ion and electron diamagnetic drifts with velocity $u_{so}=2cT_s/q_sB_oa$. Now we solve the linearized two-fluid equations. The time derivatives of the x component of (\ref{eq:momentum}), in which $\partial_t(n_sq_s)$ is eliminated using (\ref{eq:continuity}) and $\partial_tu_{sy}$ is eliminated using the y component of (\ref{eq:momentum}), are
\begin{eqnarray}
\partial_tE_{x}=4\pi\frac{v_A^2}{c^2}[J_{ix}-\rho_i^2(\partial_{xx} J_{ix}+\partial_{xz} J_{iz})+
\omega_i^{-2}\partial_{tt}J_{ix}\nonumber \\
+\frac{u_{io}}{\omega_i}(\partial_{x}J_{ix}+\partial_{z}J_{iz})-n_iq_i\frac{E_{y}c}{B_z}]+u_{io}\partial_xE_y,\label{eq:E_x}\\
J_{ex}=n_eq_e\frac{E_{y}c}{B_z}+\frac{m_ec^2\partial_tE_{x}}{m_i4\pi v_A^2}+\rho_e^2\partial_{xz} J_{ez}-\frac{u_{eo}}{\omega_e}\partial_{z} J_{ez}, \label{eq:J_ex}
\end{eqnarray}
where $v_A^2=B_z^2/4\pi n_im_i$ is the local Alfv{\'e}n speed, $\omega_s=q_sB_z/m_sc$ is the local gyrofrequency of species $s$, and $\rho_s=\sqrt{T_s/m_s\omega^2_s}$ is the local gyroradius of species $s$. In (\ref{eq:J_ex}) we have assumed $\rho_e\partial_x \ll 1$ and $\omega_e^{-1}\partial_t \ll 1$. At the current sheet center where electrons are unmagnetized, the above assumption is ambiguous and the electron pressure anisotropy may become relevant \cite{Vasyliunas1975,Hesse1999}. In the z component of the electron momentum equation, the pressure gradient dominates the inertia effect in high $\beta$ plasmas \cite{Lysak1996,Stasiewicz2000}. Taking time derivative of this equation yields
\begin{eqnarray}
\partial_tE_z=-(\partial_{zz}J_{ez}+\partial_{zx}J_{ex})T_e/(n_eq^2_e) +u_{eo}\partial_zE_y.\label{eq:E_z}
\end{eqnarray}
Eliminating $-q_in_iu_{io}B_x/c-\partial_zn_iT_i$ in the $z$ component of ion momentum equation yields 
\begin{eqnarray}
m_in_i\partial_tu_{iz}=q_in_iE_z(1+T_i/T_e)\label{eq:J_iz}
\end{eqnarray}
Substitute (\ref{eq:Ampere}), (\ref{eq:E_x}), and (\ref{eq:E_z}) into time derivative of the y component of (\ref{eq:Faraday}). Rescaling $x/a\rightarrow x$ and Fourier transforming the achieved equation in z and t ($\partial_t \sim -i\omega,\partial_z \sim ik_z$) gives 
\begin{eqnarray}
\partial_x\left(\frac{\partial_{x}\widetilde{B}_y}{{\rm sech}^2\,x}\right)+\left(\frac{\omega ^2}{k_z ^2 V_A^2}-\frac{\tanh ^2x}{{\rm sech} ^2 \,x}\right)\frac{a^2\widetilde{B}_y}{\rho_{io}^2+\rho_{ao}^2}=\frac{\widetilde{S}}{k_z^2}, \label{eq:Inhomogeneous equation}
\end{eqnarray}
where $V_A=B_o/\sqrt{4\pi n_om_i}$ is the characteristic Alfv{\'e}n speed, $\Omega_{i}=q_iB_o/m_ic$, $\rho_{io}=\sqrt{T_i/m_i\Omega^2_i}$, $\rho_{ao}=\sqrt{T_e/m_i\Omega^2_i}$ and
$\widetilde{S}(x,\omega,k_z)$ is the Fourier transform of 
\begin{eqnarray}
S=\frac{4\pi}{c}\left(\frac{2n_oq_ic}{B_o}\partial_{zx}E_y-n_oq_ia\partial_{zzx}u_{iz}+\partial_{zx}\frac{\partial_x J_{ex}}{\text{sech}^2\,x}-\frac{m_ia^2}{T_i+T_e}\partial_{ttz}J_{ex}\right). \label{eq:source}
\end{eqnarray}
In (\ref{eq:Inhomogeneous equation}) we have neglected $\omega^2/(\Omega^2_i\text{sech}^2\,x)$ compared with $(\omega ^2/k_z ^2 V_A^2-\tanh ^2x/\rm {sech} ^2 \,x)$ in coefficients of $\widetilde{B}_y$ assuming that $B_y$ is of low frequency $(\omega^2/\Omega^2_i\ll1)$ and long parallel wavelength $(k^2_zV^2_A/\Omega^2_i\ll 1)$ as indicated by multi-spacecraft measurements \cite{Wygant2005,Vaivads2004}. We also neglect terms on order of $O(m_e/m_i)$ in (\ref{eq:Inhomogeneous equation}). Setting $\widetilde{B}_y=\psi \text{sech}\,x$ we turn (\ref{eq:Inhomogeneous equation}) into 
\begin{eqnarray}
\partial_{xx} \psi+[\lambda-1-(R^2+\lambda)\tanh ^2x]\psi=\widetilde{S}{\rm sech}\,x/k_z^2, \label{eq:Inhomogeneous Schrodinger equation}\\
\textrm{where }\lambda=\frac{\omega ^2R^2}{k_z ^2 V_A^2},R=\frac{a}{\sqrt{\rho_{io}^2+\rho_{ao}^2}}.\nonumber
\end{eqnarray}
(\ref{eq:Inhomogeneous Schrodinger equation}) is an inhomogeneous Sturm-Liouville equation. $\lambda$ is an unspecified parameter. The weight function is ${\rm sech}^2x$.\\

The homogeneous form of (\ref{eq:Inhomogeneous Schrodinger equation}) is a time-independent Schrodinger equation with corresponding total energy $E=\lambda-1$ and a potential well $V\tanh^2x$, $V=R^2+\lambda$.  Only bound state solutions exist since $E<V$. The allowed energy levels (see Ref.\cite{Morse1953}, p.1653) $E=V-[\sqrt{V+\frac{1}{4}}-(n+1)]^2$ yield eigenvalues of $\lambda$ 
\begin{eqnarray}
\lambda_n=(n+1)n+(1+2n)\sqrt{R^2+1}+1. 
\end{eqnarray}
The eigenfunctions are
\begin{eqnarray}
\psi_n(x)=\frac{F(-n,2\sqrt{V(\lambda_n)+\frac{1}{4}}-n,|b_n+1|,\frac{1}{e^{2x} +1})}{(e^{x}+e^{-x})^{b_n}},
\end{eqnarray}
where $F$ is the hypergeometric function, $b_n=\sqrt{V(\lambda_n)-E(\lambda_n)}=\sqrt{R^2+1}$. Eigenfunctions are real and the first two are $\psi_0(x)=(e^{-x}+e^{x})^{-\sqrt{R^2+1}}$ and $ \psi_1(x)=(e^{-x}+e^{x})^{-\sqrt{R^2+1}}(e^{2x}-1)/(e^{2x}+1).$
The eigenmode is Alfv{\'e}nic as indicated by its phase velocity $\lambda$ introduced in (\ref{eq:Inhomogeneous Schrodinger equation}). The Harris sheet confines the Alfv{\'e}n eigenmode in the same way a $\tanh ^2(x)$ potential well confines a quantum mechanical wave. From the perspective of mode conversion theory \cite{Stix1992}, Kinetic Alfv{\'e}n Wave (KAW) provides a useful insight to the eigenmode solution. The KAW dispersion relation is $\omega ^2/(k_z ^2v_A^2)=1+k_x^2(\rho_i^2+\rho_a^2)$ \cite{Lysak1996,Stasiewicz2000}. We can heuristically achieve a similar Schrodinger equation with $\tanh^2$ potential by replacing $k_x$ with $-i\partial_x$ in the KAW dispersion relation and employing the x dependence of $v_A^2,\rho_i^2$ and $\rho_a^2$ from the Harris sheet.\\

With the eigenmode solutions we proceed to calculate $\widetilde{B}_y$ from equation (\ref{eq:Inhomogeneous Schrodinger equation}) as  
\begin{eqnarray}
\widetilde{B}_y={\rm sech}\,x\int_{-\infty}^{\infty}G(x|x_o;\omega,k_z)
{\rm sech}\,x_o\frac{\widetilde{S}(x_o,\omega,k_z)}{k_z^2}dx_o, \label{eq:B_y Fourier}
\end{eqnarray}
where the Green's function $G=\sum_n \psi_n^\ast(x_o)\psi_n(x)/[(\lambda-\lambda_n)\Phi_n^2]$ is an infinite series of eigenfunctions and $\Phi_n^2=\int_{-\infty}^{\infty}\psi_n(x)\psi_n^\ast(x){\rm sech}^2\,xdx$ is the normalization constant (see Ref.\cite{Morse1953},chapter 7). Let $\omega=\omega _r+i \eta, \eta \rightarrow 0^{+},s=\omega/i$. $s$ is the Laplace transform variable. We replace the Fourier transform in time with a Laplace transform and treat (\ref{eq:B_y Fourier}) as an initial value problem. For simplicity the initial condition is set as $B_y|_{t=0}=0,\partial_tB_y|_{t=0}=0$. The Laplace and Fourier inversion of (\ref{eq:B_y Fourier}) is 
\begin{eqnarray}
B_y=\sum_n\psi_n(x){\rm sech}\,x
\int_{-\infty}^{\infty}\int_{-\infty}^{\infty}\int_{0}^{t}
\frac{-V_A\psi_n^\ast(x_o){\rm sech}\,x_o}{2\Phi_n^2\sqrt{\lambda_n}R} \nonumber \\
\times H[(t-t_o)V_A\sqrt{\lambda_n}R^{-1}-|z-z_o|]S(x_o,t_o,z_o)dt_o dz_odx_o, \label{eq:B_y} 
\end{eqnarray}
where $H$ is the unit step function. $B_y$ is in the form of superposition of eigenmodes propagating in $\pm z$. The phase velocity of the $n$th mode is $V_A\sqrt{\lambda_n}R^{-1}$. The sources $u_{iz},J_{ex}$ and $E_y$ determine the term $S$ and thus $B_y$. Terms $J_{ix},J_{ez},E_x$ and $E_z$, grouped with $B_y$ as parts of the Alfv{\'e}n eigenmode, can be calculated from the sources using equations (\ref{eq:Ampere}), (\ref{eq:Faraday}), (\ref{eq:E_x}) and (\ref{eq:E_z}). The other half of the formulation is the response of the sources to the Alfv{\'e}n eigenmode. The field-aligned ion jet $u_{iz}$ is calculated from ~(\ref{eq:J_iz}); $J_{ex}$ is calculated from ~(\ref{eq:J_ex}); $E_y$, usually called reconnection electric field, is calculated from 
\begin{eqnarray}
\partial_{xx}\widetilde{E}_y-k_z^2a^2\widetilde{E}_y=(4\pi a^2/c^2)\partial_t\widetilde{J}_y,\label{eq:2D}\\
\partial_t\widetilde{J}_y=\frac{c^2}{4\pi a^2}\partial_x \left( \frac{\text{sech}^2x}{\tanh x}\widetilde{E}_y\right)+\frac{c^2\text{sech}^2x}{4\pi\delta_{io}^2}\widetilde{E}_y+\widetilde{S}_E, \label{eq:y momentum}
\end{eqnarray}
where (\ref{eq:2D}) and ~(\ref{eq:y momentum}) are Fourier transformed in z, $\delta_{io}=\sqrt{c^2m_i/(4\pi q^2_in_o)}$, and  $\widetilde{S}_E=u_{io}ik_z\widetilde{J}_{ez}-\omega_e\rho^2_eik_z\partial_x\widetilde{J}_{ez}-\omega_i\widetilde{J}_{ix}+(\omega_i{c^2/4\pi v_A^2})\partial_t\widetilde{E}_x$. (\ref{eq:y momentum}) is the sum of the y component of ({\ref{eq:momentum}) multiplied by $q_s/m_s$. In obtaining (\ref{eq:y momentum}) we eliminate $J_{ex}$ using ({\ref{eq:J_ex}), set $n_i=n_e$ and eliminate $\partial_t(n_eq_e)$ using (\ref{eq:continuity}). Setting $\widetilde{E}_y=\varphi\sqrt{\tanh x}$ and inserting (\ref{eq:y momentum}) into (\ref{eq:2D}) yield
\begin{eqnarray}
\partial_{xx}\varphi+[-k_z^2a^2+V(x)]\varphi=(4\pi a^2/c^2)\widetilde{S}_E/\sqrt{\tanh x}, \label{eq:Vx}
\end{eqnarray}
where $V(x)=[2\cosh(2x)-1]\text{csch}^2(2x)-a^2\text{sech}^2\,x/\delta^2_{io}$. $V(x)$ is roughly approximated as $1/(4 x^2)$. Equation (\ref{eq:Vx}) is an inhomogeneous Sturm-Liouville equation. Two independent solutions to its homogeneous form are $\varphi_1=\sqrt{x}H_0^{(1)}(-ik_zax)$ and $\varphi_2=\sqrt{x}H_0^{(2)}(-ik_zax)$ if $k_z>0$, and 
$\varphi_1=\sqrt{x}H_0^{(2)}(-ik_zax)$ and $\varphi_2=\sqrt{x}H_0^{(1)}(-ik_zax)$ if $k_z<0$. $H_0^{(1)}$ and $H_0^{(2)}$ are Hankel functions of first and second kind. We select $\varphi_1$ and $\varphi_2$ to ensure a real $E_y$, $\overline{\widetilde{E}_y(-k_z)}= \widetilde{E}_y(k_z)$. $\varphi_1|_{x\rightarrow-\infty}=0$ and $\varphi_2|_{x\rightarrow\infty}=0$ satisfy the left and right boundary conditions, respectively.  The Green's function of (\ref{eq:Vx}) is $g(x|x_o;k_z)=-\varphi_1(x_<)\varphi_2(x_>)/\Delta(\varphi_1,\varphi_2)$ (see Ref.\cite{Morse1953}, chapter 7), where $x_<(x_>)$ is the smaller (larger) of $x$ and $x_o$. $\Delta(\varphi_1,\varphi_2)=\mp4i/\pi$ is the Wronskian of $\varphi_1$ and $\varphi_2$. Using Green's function, we calculate $\widetilde{E}_y$ as the response to the Alfv{\'e}n eigenmode  
\begin{eqnarray}
\widetilde{E}_y=\sqrt{\tanh x}\int_{-\infty}^{\infty} g(x|x_o;k_z)\frac{-(4\pi a^2/c^2)}{\sqrt{\tanh x_o}}\widetilde{S}_E dx_o.\label{eq:E_y}
\end{eqnarray}
Once we obtain $E_y$ by Fourier inversion in $z$, we calculate $B_x,B^{'}_z$ and $J_y$ through (\ref{eq:Ampere}) and (\ref{eq:Faraday}). Now we complete a dynamic formulation of collisionless reconnection. The existence of parallel $E_z$ indicates the breaking of the ``frozen-in'' condition.\\ 

The physical meaning of the calculation has two reciprocal parts: The sources ($E_y,J_{ex},$ and $u_{iz}$) generate Alfv{\'e}n eigenmodes propagating outward in $\pm z$; Meanwhile the Alfv{\'e}n eigenmodes excite the sources and dissipate. The eigenmodes-sources coupling evolves self-consistently following an initial perturbation. Before the system reaches the phase in Fig.\ref{fig:fig1}, we expect an eigenmodes-generation phase in which magnetic energy is converted to establish Hall fields and Hall currents. We try to produce this phase with a test perturbation $E_y=\left|E_o\right|\exp[-(x^2+z^2)/l^2]$ that is associated with suitable $\partial_t \mathbf{B}$ in extracting magnetic energy and changing magnetic topology globally, $l$ can be as large as the system scale. We assume that reconnection ion jets are not established ($u_{iz}\approx 0$) in this phase. We also assume that electrons approximately $\mathbf{E}\times\mathbf{B}$ drift in $x$ ($J_{ex}\approx n_eq_eE_yc/B_z$) and avoid any evaluation around $x=0$. Calculation shows $S$ (\ref{eq:source}) is quadrupole, $sign(S)=-sign(xz)$. Evaluation of (\ref{eq:B_y}) shows that the n=1 mode dominates and $sign(B_y)=-sign(S)$. The contribution from n=1 mode is $B_y=C\psi_1(x)\text{sech} x \partial_z\exp[-(z/l)^2]t^2$, where $sign(\psi_1(x))=sign(x)$ and $C$ is a negative constant. In this phase the quadrupole Hall $B_y$ and Hall current are opposite to those in Fig.\ref{fig:fig1}. The Hall current $J_x \approx J_{ix}$, like the perpendicular current in a KAW, is mainly a modified ion polarization current and associated with the increasing of inward Hall $E_x$ in the region $z/l<1$. Electrons move along the magnetic field to track drift ions and keep quasi-neutral, producing the consistent parallel Hall current $J_z$, $J_z\approx J_{ez}$. In the parallel direction force balance is approximately true for electrons, $n_eq_eE_z\approx n_eq_eu_{eo}B_x/c+\partial_zn_eT_e$. The contribution to $J_zE_z$ from $n_eq_eu_{eo}B_x/c$ almost cancels out in integrating over $x$. The density gradient in $z$ created by the drift ions is inward at inner current sheet and outward on the outer periphery, relating a total $J_zE_z>0$. $J_xE_x>0$ and $J_zE_z>0$ indicate that the eigenmode stored the converted magnetic energy in the form of increasing wave energy. The time scale of this phase is the ion polarization drift time, equal to the time of establishing the Hall $E_x$. According to (\ref{eq:J_iz}) ion experiences an total outward force $q_in_iE_z(1+T_i/T_e)$ in the inner current sheet. The system probably transits to the eigenmode-dissipation phase when ion acceleration becomes dominating. The ion jets' term can dominate $S$ and produce the Hall quadrupole pattern in Fig.\ref{fig:fig1}. In this phase $J_xE_x<0$ and $J_zE_z<0$ indicate a decrease in the wave energy transfered to the accelerated ion jets and the excitation of secondary $E_y$. In the region $x/a\gg 1$, the excited $\widetilde{E}_y$ goes as $\tilde{A}(k_z)\sqrt{\tanh x}\varphi_2\sim \tilde{A}(k_z)\exp(-\left|k_z\right|x)$ according to (\ref{eq:E_y}). The outward ion jet picks up energy $q_iE_zl\approx q_iE_xa$. We notice $q_iE_xa \sim (1/2)m_iV_A^2$ in observations  \cite{Mozer2002,Wygant2005,Vaivads2004}. \\

Equation (\ref{eq:B_y}) explicitly indicates several results. In reconnection $S$ is an odd function of $x$ and eliminates all even modes. The amplitude factor $1/(\sqrt{\lambda_n}R)$ indicates that Alfv{\'e}n eigenmodes are created easier in thinner current sheet and that the n=1 mode dominates. The step function $H$ suggests that reconnection process can extend in $\pm z$ at the velocity $V_A\sqrt{\lambda_1}/R$ in this eigenmodes-sources coupling formulation. A local change can be communicated with the rest of the system over the dynamical timescale $LRV_A^{-1}\lambda_1^{-1/2}$. $L$ is the system scale. The dynamical timescale is often related to the time taken for a system to respond to a change and move to another equilibrium state. \\  

Fig.\ref{fig:fig2} presents the comparison between the n=1 Alfv{\'e}n eigenmode and the measurement of Hall fields from Polar satellite \cite{Mozer2002}. We model the measured current sheet in Ref.\cite{Mozer2002} as a Harris sheet with parameters $n_o=8cm^{-3},B_o=80nT$ and $a=150km$; these numbers are from the observation. $T_i=5T_e$ is our estimate for a typical current sheet. Independent determination of the amplitude and sign of the Hall fields needs past information, which is unavailable. Therefore we take the measured amplitude of $B_y$ as an input. We compare the x dependence of the Hall fields with the form of the n=1 mode in the present theory. The perturbation assumption ($\delta n_s/n_s \lesssim 1, B_y/B_o\lesssim 1$ and $\delta B_z/B_o\ll 1$) is roughly satisfied in this case. In a pure n=1 mode $B_y\sim\psi_1(x)\text{sech} x $ and $ E_x\sim(V_AR/c\sqrt{\lambda_1})\sinh^2x [1-(\rho_i^2/a^2)\partial_{xx}+(u_{io}/\omega_i)\partial_x]B_y$. $E_x$ is estimated with the absence of sources. $B_y$ and $E_x$ show good agreement with observations in the x dependence. A minor difference may result from the deviation of density from $n_o\text{sech}^2\,x$. In addition, given the measured amplitude of $B_y$, the calculated amplitude of $E_x$ shows good agreement with data. We also suggest searching n=3 mode signals in lab experiments \cite{Ren2005,Yamada2006}.\\ 

In conclusion, this paper addresses the most fundamental issues of reconnection, namely the energy conversion mechanism and the time scale. We propose a new mechanism of generating and dissipating Alfv{\'e}n eigenmodes for time-dependent collisionless reconnection. The dynamical timescale of reconnection, determined as the system scale $L$ divided by the eigenvalue propagation velocity $V_A\sqrt{\lambda_1}/R$, approaches the Alfv{\'e}n transit time $L/V_A$ as $R\rightarrow 1$. This can be much faster than Sweet-Parker and Petschek reconnection models. Notice that the physical meaning of the key result is different in each mechanism. Both the diffusion velocity and shock propagation velocity are local outward velocities in $\pm x$ that balance the inflow velocity at a certain interface; the eigenvalue propagation velocity in our approach is a velocity in $\pm z$ at which a local perturbation communicates globally with the rest of system. The growth rate measures the time scale for unstable reconnection modes to grow significantly; the dynamical timescale in our approach implies an interval over which a new equilibrium is achieved.\\ 

The physics of plasma heating and energization of high energy particles in reconnection are not resolved in this paper. Isothermal electrons may be an appropriate approximation since $\omega/k_zv^e_{th} \ll 1$ \cite{Stix1992}. Isothermal ions can be modified to resolve ion heating through Landau damping. A kinetic treatment or test particle method is needed to understand the formation of high energy particles.\\   

The author is grateful to R.Lysak, J.Wygant, C.Cattell, S.Thaller, L.B.Wilson III, Y.Song, J.Woodroffe, Lian Chen and Xin Tao for valuable discussions.

\end{document}